\documentstyle[emulateapj]{article}
\received{}
\revised{}
\accepted{}
\journalid{}{}
\articleid{}{}
\paperid{}
\cpright{}{}
\ccc{}
\lefthead{Matthews \& Gallagher}
\righthead{Rotation Curves of Low-Luminosity Spirals}
\begin{document}

\title{High-Resolution Optical Rotation Curves of Low-Luminosity 
Spiral Galaxies}

\vskip0.5cm
\author{L. D. Matthews\altaffilmark{1}}
\author{J. S. Gallagher, III\altaffilmark{2}}

\altaffiltext{1}{Harvard-Smithsonian Center for Astrophysics,
60 Garden Street, MS-42, Cambridge, MA 02138 USA; Electronic mail: 
lmatthew@cfa.harvard.edu}
\altaffiltext{2}{Department of Astronomy, University of Wisconsin-Madison,
475 N. Charter Street, Madison, WI 53706 USA}

\newcommand{\ang}{\rm \AA}
\def\lsim{~\rlap{$<$}{\lower 1.0ex\hbox{$\sim$}}}
\def\gsim{~\rlap{$>$}{\lower 1.0ex\hbox{$\sim$}}}
\newcommand{\nin}{\noindent}
\newcommand{\jks}{\mbox{Jy$\cdot$km s$^{-1}$}}
\newcommand{\nan}{Nan\c{c}ay}
\newcommand{\skp}{\mbox{  }\\}
\newcommand{\kms}{\mbox{km s$^{-1}$}}
\newcommand{\msun}{\mbox{${\cal M}_\odot$}}
\newcommand{\lta}{\stackrel{<}{\sim}}
\newcommand{\ewha}{W$_\lambda$(H$\alpha$)}
\newcommand{\gta}{\stackrel{>}{\sim}}
\newcommand{\ecs}{erg~cm$^{-2}$~s$^{-1}$}
\newcommand{\mhi}{\mbox{${\cal M}_{\HIss}$}}
\newcommand{\HI}{\mbox{H\,{\sc i}}}
\newcommand{\HII}{\mbox{H\,{\sc ii}}}
\newcommand{\SII}{\mbox{S\,{\sc ii}}}
\newcommand{\NII}{\mbox{N\,{\sc ii}}}
\newcommand{\HIapp}{\mbox{H\hspace{0.155 em}{\scriptsize I}}}
\newcommand{\HIbf}{\mbox{H\hspace{0.155 em}{\scriptsize \bf I}}}
\newcommand{\HIit}{\mbox{H\hspace{0.155 em}{\scriptsize \it I}}}
\newcommand{\HIss}{{\rm H\hspace{0.04 em}\scriptscriptstyle I}}
\newcommand{\darkV}{$\frac{{\cal M}_{HI}}{L_{V}}$}
\newcommand{\lsun}{\mbox{${\cal L}_\odot$}}
\newcommand{\dark}{${\cal M}_{HI}/L_{B}$}
\newcommand{\am}[2]{$#1'\,\hspace{-1.7mm}.\hspace{.0mm}#2$}
\newcommand{\as}[2]{$#1''\,\hspace{-1.7mm}.\hspace{.0mm}#2$}

\begin{abstract}
We present optical longslit spectroscopic observations of
21 low-luminosity, extreme late-type spiral
galaxies. Our sample is comprised of Sc-Sm Local Supercluster
spirals with moderate-to-low optical surface brightnesses and with
luminosities at the low end for spiral disk galaxies 
($M_{V}\ge-18.8$). For each 
galaxy we have measured
high spatial resolution position-velocity (P-V) curves 
using the H$\alpha$ emission line, and for 15 of the galaxies we also derive
major axis rotation curves.
In $\sim$50\% of our sample,
the P-V curves show significant asymmetries in shape,
extent, and/or amplitude on the approaching and receding sides of the disk.
A number of  the P-V curves are
still rising to the last measured point, or reach a clear turnover 
on only one side. In most instances we find good agreement between the 
kinematic centers of extreme late-type spirals as defined by  the 
global \HI\ emission profile and by 
their optical continuum, although in a few cases
we see evidence of possible real offsets.
In spite of their shallow central gravitational potentials,
at least 6 of the galaxies in our sample 
possess semi-stellar nuclei that appear  to be compact
nuclear star clusters; in 
5 of these cases we see 
kinematic signatures in the P-V curves at the location of the
nucleus.  Finally, 
we find that like giant spirals, our sample galaxies
have higher specific
angular momenta than predicted by current cold dark matter models.

\end{abstract}


\section{Introduction\protect\label{intro}}
\subsection{Background}
Matthews \& Gallagher (1997; hereafter MG97) used 
the term ``extreme late-type
spirals'' to refer to the lowest-luminosity rotationally-supported disk
galaxies that still exhibit regular disk structure and well-defined
optical centers. Typically these are moderate-to-low surface
brightness Sc-Sm spirals that represent the  low-end extremes of
properties such as mass, luminosity, and size for a given Hubble type.
 
Extreme late-type spirals are among the 
dynamically simplest of disk galaxies: they tend to be pure disk
systems, often with weak or absent spiral arm structure. 
Some extreme late-type spirals also
appear to have undergone only minimal dynamical heating (e.g.,
Matthews 2000). However, in spite of their
apparent simplicity, there is evidence  of 
a variety of intriguing structural and
dynamical phenomena in these small galaxies,
including: multi-component disk structure (Matthews 2000);
kinematically lopsided disks (e.g., Matthews, van Driel, \&
Gallagher 1998; Swaters et al. 1999; Matthews \& Uson 2002); and
compact starcluster 
nuclei (e.g., Matthews et al. 1999; B\"oker et al. 2001,2002). High-quality
kinematic information for extreme late-type spirals is clearly
desirable in order to further explore these trends. 

Recently, the acquisition of 
kinematic information, particularly disk rotation curves, 
for low-luminosity and low surface brightness (LSB) spirals has also
become of considerable 
interest for exploring broader issues relevant to our overall
understanding of galaxy formation and the nature of galaxian
dark matter.
From numerous studies, it has now been found that 
the dynamics of these types of galaxies
often appear to be dominated by dark matter even at
very small galactocentric
radii (e.g., Jobin \& Carignan 1990;
C\^ot\'e, Carignan, \& Sancisi 1991; Martimbeau, Carignan, \& Roy
1994; Meurer et al. 1996; McGaugh \& de Blok 1998; de Blok et al. 2001). 
Because the contribution to the 
overall dynamics from
the visible matter in these faint disk systems is consequently quite small, 
this means that
extreme late-type/LSB spirals can provide us with some of the most stringent
constraints on dark matter halo properties of galaxies, and may hold a
key to testing important classes of galaxy formation paradigms and cosmological
models (e.g., Kravtsov et al. 1998; van den Bosch et al. 2000;
Iliev \& Shapiro 2001; van den Bosch, Burkert, \& Swaters 2001). 

Until recently, most of the  rotation curves that have been obtained for
extreme late-type spirals, especially LSB systems, were derived
from \HI\ aperture synthesis observations (e.g., Carignan, 
Sancisi, \& van Albada 1988; Bosma, van der Hulst,
\& Athanassoula 1988; van der Hulst et
al. 1993; de Blok, McGaugh, \& van der Hulst  1996; 
Swaters 1999).  
Unfortunately, many of the \HI\ rotation curves available for
extreme late-type spirals suffer from
poor angular resolution, leading in some cases to significant beam smearing, 
whereby the slope of the inner,
rising portion of the rotation curve is underestimated 
(see Swaters 1999; Swaters, Madore, \& Trewhella 2000; van den
Bosch et al. 2000). This in turn
hampers attempts to perform accurate rotation curve mass
decompositions and to meaningfully
constrain the shape of the dark matter halo (e.g., Blais-Ouellette et
al. 1999;
van den Bosch et al. 2000).  
Beam smearing can be particularly problematic for the extreme late-type
spirals, since these systems 
are physically small, and at distances of 10-20~Mpc, have angular
sizes of only a few arcminutes or less. 
For this reason, {\sl optical}  kinematic data
(i.e., longslit emission line spectroscopy or
Fabry-Perot interferometry), which have higher spatial and
spectral resolution, offer a more accurate means to map the inner
rotation curve of extreme late-type spirals. These observations can
thus serve as an extremely
valuable complement to \HI\ observations in mapping the full disk
rotation curves (e.g., de Blok et al. 2001). The combination of
optical and \HI\ data is also powerful for constraining disk angular
momenta (e.g., van den Bosch et al. 2001; see also 
Section~\ref{angmom}) and for 
comparing the
kinematics of the ionized versus the neutral gas. 
Finally, high-resolution optical
kinematic data provide unique information in their own right, such as 
kinematically
confirming the presence of compact star cluster nuclei in small spirals
(see Section~\ref{nukes}).

\subsection{The Present Study\protect\label{present}}
MG97 presented optical $B$ and $V$ CCD
photometry for a sample of 49  extreme late-type spiral galaxies with
$\delta< -18^{\circ}$.
Their sample consisted of Sc-Sm field spirals, all within the 
Local Supercluster ($V_{h}\le$3000~\kms.). 
Although these objects are of relatively low luminosity and span a variety
of morphological types (see Figure~11 of MG97), all have regular
disk structures and well-defined optical centers, distinguishing them
structurally from dwarf irregular (dIrr) galaxies of
similar luminosities.
Most, but not all of the galaxies are objects of low mean optical
surface brightness ($\bar\mu_{V,i}\gsim23$~mag~arcsec$^{-2}$).
Generally these galaxies are \HI-rich, with high \dark\ ratios for their
Hubble types (MG97; Matthews, van Driel, \& Gallagher 1998). Finally,
all of the galaxies in the MG97 sample
exhibit significant rotational broadening in their
global \HI\ profiles, with all but few objects showing 
double-peaked \HI\ spectra (Fouqu\'e et al. 1990;
Matthews et al. 1998).

In this paper we use optical longslit emission line (H$\alpha$)
spectroscopy to explore the rotational properties of 21 of the MG97
galaxies. In Section~\ref{sample} 
we describe our sample selection. In Section~\ref{obs}
\& \ref{reduction} 
we describe our observations and data reduction, respectively.  
We present position-velocity (P-V) diagrams derived
from the new H$\alpha$ data in Section~\ref{results},
and compare them to the global \HI\
profiles of the galaxies. For the 15 galaxies with highest-quality
optical data, we also derive
major axis rotation curves. 
We discuss characteristics of our P-V
plots and rotation curves further in Section~\ref{results}, 
including the presence
of kinematic asymmetries (Section~\ref{asymm}).
In Section~\ref{nukes}, 
we investigate the dynamical signatures of the semi-stellar
nuclei on the rotation curves of several of the galaxies and use these
to derive crude mass estimates for the nuclei, while 
in Section~\ref{angmom} we briefly comment on the specific angular momenta of
the disks in our sample.
Finally, in the Appendix we provide comments on our 
individual longslit spectra. 

Our optical disk
rotation profiles are measured using
the H$\alpha$ emission line, which is in general only easily observable to
within $r\lsim D_{25}$ in most normal, bright galaxies, and to only a
fraction of $D_{25}$ in faint, extreme
late-type systems (e.g., Table~2). 
Therefore, ultimately the observations we present should be
combined with extended \HI\ rotation curves to gauge the full disk
rotation of our sample galaxies. Currently \HI\ rotation curve data
are not available for the bulk of the galaxies in our sample, hence
it is intended that the results presented here will provide an aid 
for selecting objects suitable for future \HI\ mapping.
For this reason, we  are also making available in electronic format
all of our measured P-V data (Table~3).

\section{Sample Selection\protect\label{sample}}
\begin{figure}
\plotfiddle{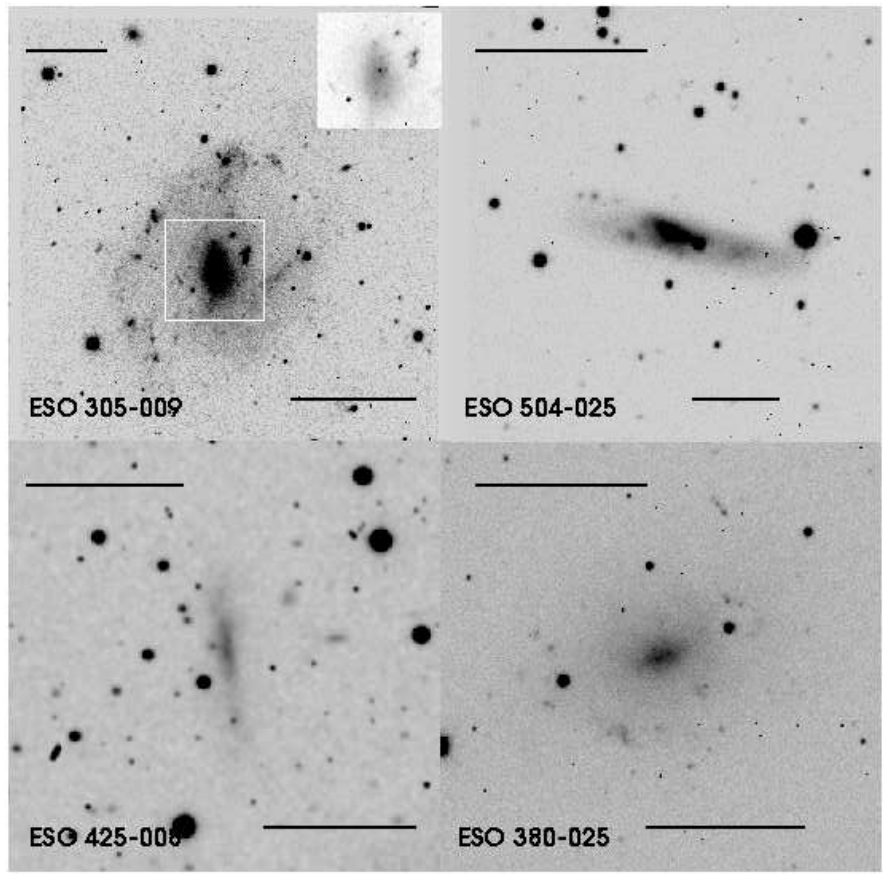}{3.0in}{-90}{60}{60}{-230}{400}
\caption{Sample $V$-band CCD images of 4 of our program
galaxies: ESO~305-009 (top left), ESO~504-025 
(top right), ESO~425-008 (bottom left), 
and ESO~380-025 (bottom right). The scale bars on each panel indicate
1 arcminute (upper) and 5 kpc at the adopted distance of the galaxy
(lower). The inset on the ESO~305-009
panel highlights the pointlike nucleus in this galaxy. Narrow band
H$\alpha$ images as well as our new H$\alpha$ spectroscopy confirm
this feature is not a foreground star (see Section~\ref{nukes}). ESO~380-025
also has a pointlike nucleus. These imaging data are 
described in detail in MG97.\protect\label{fig:elts}}
\end{figure}

For our spectroscopic measurements
we selected from MG97 
twenty-one extreme late-type spiral galaxies covering a range of optical
morphologies.  
Where possible, preference was given to
galaxies with moderate-to-high inclinations.  All of our targets have
radial velocities $V_{h}<$2800~\kms\ and therefore reasonable 
spatial resolution for these systems can be
achieved from the ground under average seeing
conditions. Some global properties of our
targets are given in Tables~1 \& 2. Sample broadband 
images of 4 of the galaxies are shown in Figure~\ref{fig:elts}.

\section{Observations\protect\label{obs}}
Our longslit spectroscopic observations were obtained during two nights
in 1995 January using the 4-m telescope and Ritchey-Chr\'etien
(R-C) Spectrograph
at the Cerro Tololo Inter-American Observatory (CTIO)\footnote{Cerro
Tololo Inter-American Observatory is operated by the Association of
Universities for Research, Inc. under contract with the National
Science Foundation}.
All spectra were taken under clear conditions.
Seeing throughout the run was $\sim$\as{0}{8}-\as{1}{0}.

Our R-C spectrograph set-up included the folded Schmidt camera at
the  {\it f}/7.8 focus. The detector was a Tek 1024$\times$1024 CCD
with  Arcon controller. With the Arcon controller, the CCD
is divided into four quadrants, each with its own amplifier. This
results in a slightly different gain and readnoise
for each of the four sections of
the chip. The mean gain was 1.25~$e^{-}$/ADU and the mean readnoise
was 4.3~$e^{-}$ rms. At the  {\it f}/7.8 focus, the folded Schmidt camera
yielded a plate scale
of \as{0}{8} per 24$\mu$m pixel on the Tek CCD. Our slit
length and width were  328$''$ and \as{1}{5}, respectively. We used
 filter GG495
with grating
KPGLG, which has 860~{\sl l}/mm, a blaze
of 11000\ang\ at first order.
This yielded  a
dispersion of 0.68\ang\ per pixel and
 spectral coverage $\Delta\lambda$ of roughly 615\ang\ centered on the
redshifted H$\alpha$ $\lambda\lambda$6562.82\ang\ line.
This $\Delta\lambda$ interval also includes the
[\NII] $\lambda$6583.41\ang, and [\SII]
$\lambda\lambda$6718.26,6732.64\ang\ emission lines.

Using position angles from the literature or 
estimated from previously obtained CCD images,
we generally strove to place the spectrograph
slit along the major axis of each target (or in two cases, along the
galaxy bar; see Table~1).  Unfortunately,
in a few cases, less optimal slit position angles were
achieved 
(see Table~1 and the Appendix). Hereafter we denote the offset in 
angle between the photometric major axis and the observed slit 
position angle as $\Delta\phi$.  All of our target
galaxies have small enough optical angular sizes such that they
were fully sampled with the 328$''$ slit length.

Our exposure times for each target
ranged from 500-2000$s$ (see Table~1). Whenever possible, two
independent exposures were obtained to aid in cosmic ray rejection. 
During each observation, a ThAr
comparison lamp exposure was also 
acquired at the same telescope position as the
target galaxy for use in correction for geometric distortions (see  below).

\section{Data reduction\protect\label{reduction}}
Our data were reduced using the standard spectral
reduction and analysis packages
available in {\it IRAF}\footnote{IRAF is distributed by the National Optical
Astronomy Observatories, which is operated by the Associated
Universities for Research in Astronomy, Inc., under cooperative
agreement with the National Science Foundation.}.
Overscan corrections were applied to each frame
using the QUADPROC routine.
Sequences of bias frames taken each afternoon showed
the bias level of the Tek CCD to be extremely unstable. Count levels
in the bias images
ranged from 3 to 8 ADU, but the underlying structure varied widely
for frames obtained only minutes apart. In addition, a ripple pattern with
an amplitude $\sim$0.5~ADU was visible in most bias
frames, and this pattern
was found to shift in location from one bias frame to another. 
For this reason,
we were not able to reliably
subtract the bias level from our images. Since the count
levels in the bias  can be significant relative to some of the faint
emission line signals from our targets, we were unable to obtain
accurate absolute
line fluxes
or line
ratios from the present data. This also introduces an additional source of
error in determining the centroid of the galaxy in cases where the
galaxy continuum is fairly weak (see below).
However, our absolute wavelength calibration
should not
be affected.

\begin{figure}
\plotfiddle{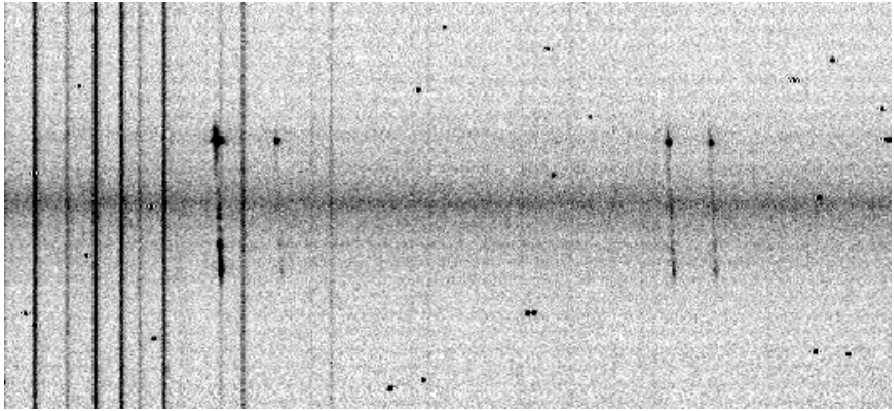}{3.0in}{0}{90}{90}{-260}{-230}
\caption{Sample longslit spectrum for
the galaxy ESO~502-016. The $x$-axis is the
dispersion axis, with $\lambda$ increasing from left to right, and the
$y$-axis is the spatial axis. The spectrum has been trimmed to show a
field-of-view along the spatial axis of $\sim$\am{2}{6}. The
horizontal band through the center of the image is the galaxy
continuum, and the bright vertical lines extending the full height of
the image are the telluric OH sky
lines. The visible galactic emission features correspond, from left to
right, to the redshifted lines of:
H$\alpha$ $\lambda\lambda$6562.82\ang\, [\NII] $\lambda$6583.41\ang\,
and [\SII] $\lambda\lambda$6718.26,6732.64\ang. The small black specks
on the image are due to cosmic rays.
\protect\label{fig:image}}
\end{figure}

The {\it IRAF} BACKGROUND task  was used to remove scattered light from our
images, and the data were 
flatfielded using normalized domeflats. The
task ILLUM was then used to apply
an illumination correction derived from twilight sky exposures. This
procedure yielded images that were flat to a fraction of a per cent. A
sample of one of our flatfielded spectra is shown in Figure~\ref{fig:image}.

Cosmic rays were identified by blinking the multiple object exposures
when available, otherwise by visual
inspection. In cases where two object exposures were available, cosmic
ray events
in the vicinity of galaxian emission features
were removed by replacing the affected pixels with the
corresponding pixel
values from the other image. When only one exposure was available,
cosmic rays were removed
by interpolating from the surrounding
pixels (in cases where they landed on background regions), or by
simply masking the affected pixel(s) (in instances where they
were superimposed on an emission line).

To derive a correction for geometric distortions, we utilized
our ThAr comparison
spectra and followed a procedure similar to that outlined by Massey,
Valdes, \& Barnes (1992). We identified the lines in the ThAr lamp 
spectra, and then applied
to our galaxy spectra the
geometric transformation of $\lambda$ as a function of the position on
the CCD by fitting a sixth order, two-dimensional Chebyshev
function using the task
FITCOORDS.
Based on the dispersion solutions found by
FITCOORDS, we transformed all rows along our object spectra to
identical linear wavelength scales using the task TRANSFORM, which
performed the necessary interpolations using a third order spline function.
The rms residuals for the wavelength solutions were typically 0.05
pixel, corresponding to 0.034~\ang\ ($\sim$1.6~\kms\ at H$\alpha$). As
a check, we
also attempted the corrections for geometric distortions using lines
identified from night sky spectra. We found the results of the two
approaches to be indistinguishable. Aside from the expected
flexure shifts (which affect absolute, but not relative wavelength
calibrations; see below), we found our
dispersion solutions to be very stable over the course of the run.

During our reductions we found the
multiple
standard star exposures obtained for the purpose of correcting the tilt
of the spatial axis relative to the edge of the detector
to be inadequate to map the trace along the full extent of the slit,
hence
we were unable to properly correct for this effect. The tilt is
approximately 2 pixels across the full 1024-pixel
width of the CCD. This does not
affect our absolute wavelength calibration, but means that when a
spectrum is extracted parallel to the edge of the CCD, the H$\alpha$,
[\NII], and [\SII] emission lines will all sample slightly different
parts of the galaxy rotation curve.  This effect matters little
for our purposes, since the
[\NII] and [\SII] emission lines were often quite weak in our
galaxies and were generally not useful for mapping the galaxy rotation
curve over significant angular extents. In addition, these lines were
frequently too noisy to
provide a reliable check on the radial velocitues derived from the H$\alpha$ features.

To derive P-V curves for our galaxies, we used
the {\it IRAF} APALL task to extract a
series of spectra along the slit for each object. Because most of the
galaxies have at least some degree of asymmetry, and
because many of the rotation profiles are  rising to the last
measured point, we were unable to use a point of symmetry to determine the
systemic velocity. However, all of the galaxies
had sufficient continuum emission such that the peak of the continuum
could be used for the determination of
the galaxy center. Galaxy centroid determination
was therefore accomplished by plotting the average of 10
columns on either side of the galaxy H$\alpha$ emission, and adopting
as the center the position of the peak value of the continuum. We
estimate this method yields a galaxy center good to roughly one-third
of a pixel (i.e. $\sim$10~\kms), accounting for the bias level uncertainties
described above.

We measured the spatial profile width of a
standard star to be roughly 10 pixels, and adopted this as our
extraction aperture width.  For each extraction, the H$\alpha$ line's
centroid was determined by averaging anywhere between 8 and 19 dispersion
lines, depending on the emission line strength and velocity dispersion
in a given object.
A series of extraction  apertures
was then placed by hand along the full extent of the slit,
with roughly 3 pixels overlap between consecutive apertures. Background
regions for each aperture were set interactively, and subtracted by fitting
a second order Chebyshev function over the region.

A by-product of the APALL task for each extracted object spectrum
is the spectrum of the subtracted sky background, which contains numerous
OH night sky lines. Because of our moderately long exposures,
these lines were of high signal-to-noise.
Osterbrock et al. (1996) compiled
an accurate list of reference wavelengths for  these lines, thus we
were able to utilize them to 
obtain a final dispersion solution for our spectra. 
Use of this wavelength calibration technique
eliminates the need to separately correct the zero-point solution
for the effects of spectrograph flexure.

As a final step,  we measured
the wavelengths of the emission lines in the extracted
spectra using SPLOT to perform Gaussian fits to the line
profiles. 
The resulting wavelengths at each point were converted to radial
velocities and corrected for heliocentric motion. 
The uncertainties in each data
point ($\pm\sigma_{V}$) were computed as $\sigma_{V}=RS^{-0.5}$~\kms, where
$R$ is the velocity resolution per pixel and $S$ is the mean signal-to-noise
ratio of the three brightest pixels in each line profile. The sizes
of the error bars plotted on Figure~\ref{fig:pv} (discussed below)
thus give an indication of the relative
strength of the H$\alpha$ line in our various target galaxies.

\section{Results\label{results}}

\subsection{Optical Versus Radio Systemic Velocities\protect\label{voptcomp}}
\begin{figure}
\plotfiddle{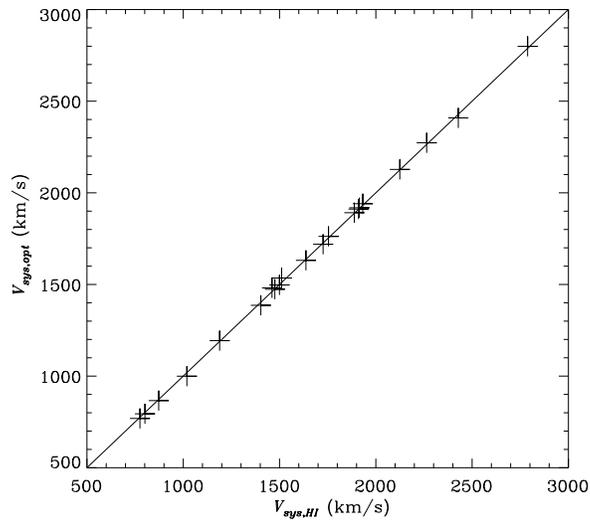}{5.0in}{0}{60}{60}{-150}{-100}
\vspace{-2cm}
\caption{Comparison of systemic radial velocities derived from our new
H$\alpha$ spectroscopic observations with those derived from 
global \HI\ measurements (see also Table~2).
\label{fig:comp}}
\end{figure}

Figure~\ref{fig:comp} shows a plot comparing the heliocentric radial 
velocities derived
from our new optical rotation curves with the values derived from
global \HI\ spectra by Fouqu\'e et al. (1990) and Matthews et
al. (1998). 
Overall the agreement is excellent. This agreement
suggests that most extreme late-type spirals seem to have 
well-defined centers in spite of their often diffuse stellar disks
and shallow central potentials. Consequently, this can offer 
an additional means of distinguishing extreme late-type
spirals from true irregular galaxies. 

In our sample we have only 6 instances where we
see a discrepancy of more than $\sim$10~\kms\ between the 
radial velocities derived from optical versus radio measurements. 
Because of the small formal measurement errors 
on both the optical and \HI\ radial velocities 
($\lsim$10~\kms\ and $\lsim$5~\kms, respectively), it is possible that in
at least some of these cases, these discrepencies may a consequence of 
true offsets between the centroid of the global \HI\ profile and the
center of the galaxy defined by the optical continuum. 
Similar offsets have been reported in other moderate-to-low-mass, 
late-type spirals (Colin \& Athanassoula
1981; 
Minniti, Olszewski, \& Reike 1993; Rownd, Dickey, \& Helou 1994). 

In 4 of the 6 galaxies where $V_{sys,opt}$ and
$V_{sys,HI}$ differ by more than 10~\kms, asymmetries are
observed in the global \HI\ profiles (see Figure~\ref{fig:pv}; 
Section~\ref{asymm}).
Together these two trends may hint that these could be
galaxies where the stellar disk is in libration about the minimum of
the \HI\ or dark matter potential (Kornreich et al. 1998; see also
Levine \& Sparke 1998; Noordermeer, Sparke, \& Levine 2001). Levine \& Sparke (1998) 
and Noordermeer et al. (2001) have shown through
numerical simulations and modelling that such configurations can reproduce the
lopsidedness observed in many extreme late-type spirals (see also below).
Nonetheless, as we discuss below, approximately one-third of our sample appears 
to exhibit kinematic lopsideness, hence in the majority of 
lopsided galaxies, the peak of the stellar density  
correlates surprisingly well with the centroid of the global
\HI\ profile.\footnote{The \HI\ systemic velocity we refer to is defined
as the midpoint of the global profile measured at a level equal
to either 20\% or 50\% of the peak flux density (Fouqu\'e et al.
1990; Matthews et al. 1998).}  The centroid of the global \HI\ profile thus 
generally seems to
be a good indicator of the location of the
minimum of the disk potential, even in most asymmetric galaxies. This would
seem to argue that the overall potential of the galaxies (established
mainly by the dark matter halo in the small systems considered here)
are symmetric on large scales,
even in cases where asymmetries in the visible matter or
velocity field are observed
(cf. Rix \& Zaritsky 1995; Jog 1999).

\subsection{Position-Velocity Plots and Rotation Curves\protect\label{curves}}
The position-velocity (P-V) 
curves derived from our final spectra are presented in
Figure~\ref{fig:pv}. 
When available, we plot next to the H$\alpha$ P-V curves the
corresponding high resolution global
\HI\ spectra from Matthews et al. (1998).
Parameters measured
from the optical 
data, including systemic velocities and the peak measured rotational
velocities, are presented in Table~2 along with photometric and global
\HI\ parameters from the literature. 

From Figure~\ref{fig:pv} and Table~2
it can be seen that among the galaxies where $|\Delta\phi|<30^{\circ}$ 
(i.e., cases where the galaxy's major axis and the spectrograph slit were 
well-aligned), the disk rotation velocities measured in the plane of
the sky from our new optical data ($\Delta V_{H\alpha}$) agree to
within measurement errors with those derived from global \HI\ data
($W_{20}$) about half 
of the time. In the remaining cases, \HII\ regions
were not observable to sufficiently large radii to reach the peak
of the rotation curves in our longslit measurements of these optically
faint galaxies.

For the 15 galaxies in our sample with $i\ge40^{\circ}$ and 
$|\Delta\phi|<30^{\circ}$, 
we derived deprojected rotation curves along the disk major axis
(Figure~\ref{fig:rc}). The disks were assumed to  
be in circular rotation, and 
the measured heliocentric radial velocities at each point were
projected onto the plane of the galaxy using the relations given by
Rubin, Ford, \& Thonnard (1980) to correct for disk inclination and 
small position angle
misalignments. In cases 
where the systemic velocities derived from global \HI\ spectra
differed from our new optically derived values, we used whichever value
minimized the asymmetry of the folded rotation curve.  
Distances for the galaxies
were adopted from MG97.

\subsubsection{The Shapes of Extreme 
Late-Type Spiral Rotation Profiles\protect\label{shapes}}
We see from Figures~\ref{fig:pv} \& \ref{fig:rc}
that most of our sample galaxies
show a  leisurely rise in rotational velocity with increasing radius,
and many have rotation profiles that 
continue to rise to the last measured point. 
Similar rotation curve shapes are frequently seen among  the
extreme late-type spiral rotation curves measured by other
workers using the H$\alpha$ emission line
(e.g., Goad \& Roberts 1981; Karachentsev 1991; Makarov,
Burenkov, \& Tyurina 1999;
Swaters 1999; Dalcanton \& Bernstein 2000; McGaugh, Rubin, \& de Blok 
2001) and
have become well-established as the hallmark of small, dark matter-dominated
disk galaxies. 
These are in stark contrast to giant spirals, where 
rotation curves typically rise rapidly to $V_{max}$ within the inner portion
of the stellar disk (cf. Casertano \& van Gorkom 1991). 

In
3 of the most 
extreme cases (ESO~418-008; ESO~504-017; ESO~440-049), we can partially 
attribute the shallow velocity gradients in
our observed P-V plots to 
position angle misalignments between the spectrograph slit and the galaxy
major axis
($|\Delta\phi|>30^{\circ}$; see Table~1). However, in all other 
instances, the slow rise of the rotation velocity as a function of
$r$  appears to be  real, and
thus be indicative of low matter densities within the inner disks of these
galaxies. For example, ESO~358-060 has a particularly shallow
rotation curve in spite of being nearly edge-on ($i$=85$^{\circ}$).
This galaxy has a very low deprojected central surface brightness
[$\mu_{V}(0)\approx$25.5~mag~arcsec$^{-2}$] and would likely be nearly
invisible if seen close to face-on in an optical broadband survey image. 
As demonstrated by Matthews \&
Wood (2001), such a shallow rotation curve cannot be explained solely by 
projection effects or internal
extinction, even in edge-on systems.

In only a few of our sample galaxies do we see  a clear
turnover or flattening of the optical P-V curve on both the
approaching and receding sides:
ESO~482-005, ESO~504-025, ESO~440-049, ESO~443-079. These
galaxies are some of the ``earlier'' Hubble type objects in our
sample (Sc-Sd), and include some of the sample galaxies 
with the most clearly-defined spiral
structure.  In some other cases, we see an apparent turnover
on only one
side of the P-V curve (ESO358-020;
ESO~425-008; ESO~438-005; ESO~380-025; ESO~508-034; ESO~444-033; see
also Section~\ref{asymm}).

We find that
even in instances where a turnover or flattening is seen, our observed
P-V curves rarely extend significantly beyond this turnover. 
Moreover, while the Sc-Scd galaxies in our sample seem
more prone to exhibiting flattening
of their P-V curves, we otherwise see no absolute correlation between
Hubble type, galaxy morphology, or luminosity
and the P-V or rotation curve shape in our sample.
For example, ESO~504-025 and ESO~358-020 both
have $M_{V}\sim$-17.5 and $i\sim 40^{\circ}$, but have very different
inner rotation curve shapes and amplitudes.  At the same time,
we also see
cases where optically rather dissimilar galaxies (e.g., ESO~380-025 and
ESO~508-034) have similar rotation
curve shapes and amplitudes. These rotation curves signatures may hold
important clues as to the evolutionary histories of these 
galaxies---for example,
the possibility that morphological tranformations may occur in 
late-type disk galaxies as a result of viscous evolution, interactions, or
tidally-induced starbursts (e.g., Gallagher \& Matthews 2001; Noguchi
2001; Matthews \& 
Uson 2002).

Rotation curves that do not flatten within the optical galaxy imply
that the bulk of the stellar disk is in solid body rotation. It has
long been known that such rotation curves are common in gas-rich
dIrr galaxies  
(e.g., Carignan \& Freeman 1988; Carignan \& Puche 1990;
C\^ot\'e et al. 1991; Casertano \& van Gorkom 1991; C\^ot\'e, Carignan, \&
Freeman 2000).
However, while most such galaxies are still
predominantly rotationally supported, many of the
dIrrs with apparent solid-body 
rotation curves also have single-peaked global \HI\
spectra, implying that solid body rotation persists to the outermost
observable regions of the  gas disk (e.g., Skillman 1996).
In contrast, most of the  galaxies in our sample
have {\sl double-peaked} global \HI\ profiles (Figure~\ref{fig:pv}; 
see also Fouqu\'e
et al. 1990; Matthews et al. 1998), which in turn implies that
the outer rotation curve contains a
flat or relatively flat region (cf. Giovanelli \& Haynes 1988), 
although it may lie outside the
stellar disk. \HI\ data will clearly be needed to trace this portion  of
the disk rotation profiles in our present sample.

\subsubsection{Kinematic Asymmetries in Extreme 
Late-Type Spirals\protect\label{asymm}}
It is evident from Figures~\ref{fig:pv} \& \ref{fig:rc} 
that asymmetries are the rule rather than the
exception for our extreme late-type spiral rotation profiles. Many of the
P-V plots and rotation curves 
show significant differences in their radial extents
on the approaching and receding sides. Often, the amplitudes of
the two sides of the rotation curves also differ (sometimes by
tens of kilometers per second), and in some
cases, the rotation curves show differing velocity gradients on the two
sides of the galaxy. Although there is frequently some uncertainty in
defining the center of any given rotation curve, our sample
nonetheless appears to contain some galaxies where the velocity
differences between the approaching and receding sides are greater
than 20\%, in contrast to Swaters et al. (1999), who suggested
that perhaps such significant (kinematic)
asymmetries may not exist.

Based on statistical studies of high-quality global \HI\ data,
Richter \& Sancisi (1994) and Haynes et
al. (1998) found that \HI\ profile asymmetries occur in approximately
50\% of normal spirals. Matthews et al. (1998) explored the frequency
of \HI\ profile asymmetries in extreme late-type spirals and 
found that the frequency of 
lopsidedness may be even higher among small late-type
spirals than among giant galaxies. Unfortunately, from asymmetries in
the global \HI\ profiles alone, it is not possible
to assess whether these asymmetries arise from asymmetric gas
distributions (e.g. as a result of accretion events; Rix \& Zaritsky
1995; Zaritsky \& Rix 1997) or from true kinematic lopsidedness, which
in turn may indicate lopsidedness in the overall galaxy potential
(e.g., Schoenmakers, Franx, \& de Zeeuw  
1997; Levine \& Sparke 1998; Jog 1999). Our
new optical spectra allow us to obtain additional insight into
this question.

Six of the 17 galaxies in the present sample for which Matthews et al.
(1998)
obtained high precision global \HI\ profiles were catagorized by
Matthews et al.
as having strong \HI\ asymmetries (ESO~358-020, ESO~422-005; ESO~502-016;
ESO~504-017; ESO~380-025; ESO~444-033). In 5 out of 6 of
these cases, we find the
optical rotation curve to have a greater amplitude on the same side as
the \HI\ ``excess'' in the global \HI\ profile. Several of these cases also
exhibit different velocity gradients on the approaching and receding
sides of the disk. This latter effect is particularly pronounced
in ESO~422-005 and ESO~444-033.
This suggests that these asymmetries
result from true lopsidedness in the disk kinematics
rather than non-symmetric matter densities 
(see Swaters et al. 1999; Noordermeer
et al. 2001).
Among the 11 remaining galaxies in our
sample whose \HI\ profiles were analyzed by Matthews et al. and not
classified as strongly asymmetric, we see a
significant rotation curve
{\sl amplitude} asymmetry in only one other case (ESO~358-015), 
and a significant
difference in the rates at which the
two sides of the rotation curve rise in only ESO~358-015 and
ESO~359-029.  Strong asymmetries in the
global \HI\ profiles therefore appear to be
good, although not perfect, predictors of
lopsidedness in the disk kinematics. As noted above, we see additional
galaxies in our sample whose rotation curves show asymmetries in terms
of their {\it
extents} on the two sides of the disk (e.g., ESO~305-009, ESO~425-008,
ESO~438-005).  However, in these latter cases, it is possible that
these asymmetries arise from asymmetric matter distributions as
opposed to true lopsided kinematics.

We conclude that the results from our small sample
are consistent with  at least
one-third of extreme late-type spirals exhibiting true kinematic
asymmetries (see also Swaters et al. 1999), 
and hence possibly asymmetries in their overall 
potentials (but see Section~\ref{voptcomp}) and with approximately
50\% of extreme late-type spirals exhibiting either kinematic {\it or}
structural asymmetries.  
Since all of the objects in our sample are
field spirals, generally lying on the peripheries of loose
groups or associations
(Gallagher, Littleton, \& Matthews 1995), it would seem unlikely that
the lopsided natures of all of 
our sample galaxies are exclusively the result of very recent 
interactions with other galaxies, although there may be some 
exceptions (e.g., Matthews
\& Uson 2002). 

Although asymmetries are common among disk
galaxies covering a wide range of masses and luminosities (e.g.,
Baldwin, Lynden-Bell, \& Sancisi 1980; Richter \& Sancisi 1994; Haynes
et al. 1998), our new data, and the findings of Matthews et al. (1998),
hint that they are equally, or perhaps more common in low-mass spirals.
Indeed, based on the models of Noordermeer et al. (2001),
pronounced asymmetries are more persistent 
in galaxy systems where the dark matter halo strongly
dominates the potential, as is expected in typical extreme late-type
spirals
(see also Levine \& Sparke 1998). 
Using models with a disk lying off-center in a dark halo,
Noordermeer et al. produced model rotation curves whose shapes 
were very similar to some of those seen in the present sample, 
i.e., showing differing slopes on the approaching and
receding sides, and flattening of the rotation  profile on  one
side only. Swaters et al. (1999) noted similar signatures in the rotation
curves of the two asymmetric extreme 
late-type spirals that they analyzed.

\subsection{Kinematic Signatures of Semi-Stellar Nuclei in Extreme
Late-Type Spirals\protect\label{nukes}}
Five of the extreme late-type spirals in the present sample were noted
by MG97
to possess a compact, semi-stellar nucleus
(ESO~359-029; ESO~358-015; ESO~305-009; ESO~504-025; ESO~505-013; see
also Figure~\ref{fig:elts}). The
nucleus of ESO~359-029 was subsequently observed with the Planetary
Camera 2 on the {\it Hubble Space Telescope} 
and demonstrated by Matthews et al. (1999) to be a true compact star cluster
nucleus analogous to those at the centers of many brighter
galaxies. Subsequently,
one additional galaxy in our sample (ESO~418-008) has also been found
to possess a compact star cluster nucleus based on {\it HST} imaging
(B\"oker et al. 2002; Windhorst et al. 2002).
Although the formation mechanism for these compact nuclei at the
centers of such diffuse galaxies with shallow central potentials
remains a mystery, these types of ``naked'' nuclei are now known to be
commonplace at the centers of late-type, pure disk spirals 
(see also van den Bergh
1995; Phillips et al. 1996; B\"oker et al. 2001,2002).

In our present sample, in
five of the cases where the presence of a  compact nuclear feature is seen
in the CCD images of MG97, 
we also see a signature of this nucleus in our spectroscopic 
data---i.e. the P-V curves
exhibit reversals, breaks, or disturbances
near the location of the nucleus (see Figure~\ref{fig:pv}). 
This strongly suggests that in all of these cases,
these nuclei are compact, massive star clusters rather than simply
small nuclear \HII\ regions. 

If we assume that the ionized gas surrounding these nuclei is in
Keplerian rotation, we can make very rough estimates of their masses 
from our present data---i.e.,
$M_{nuke}=V^{2}_{max,s}rG^{-1}$, where $G$ is the gravitational
constant, $r$ is the galactocentric distance along the major axis,
and $V_{max,s}$ is the orbital
semi-amplitude of the material orbiting the nucleus in the 
reference frame of the disk, after correction for inclination. Because
the velocity  and spatial 
resolution of our data are rather coarse relative to the
amplitudes and spatial extents of these
central rotation curve features, and because it is unclear whether the
nuclear disk material will share the same inclination as the main disk
of the galaxy, these calculations are necessarily very
crude. Nonetheless, we estimate nuclear masses 
$M_{nuke}\sim10^{6}-10^{7}$~\msun, comparable to the masses of 
low-luminosity star cluster nuclei derived for 
other late-type spirals [e.g., M33:
$M_{N}\sim2\times10^{6}$~\msun (Kormendy \& McClure 1993); 
NGC~4449: $M_{N}>4\times10^{5}$~\msun\ (B\"oker et al. 2001);
IC~342: $M_{N}\sim6\times10^{6}$~\msun\ (B\"oker, van der Marel, \& Vacca 
1999)].

As a final note, we draw attention to the fact that
some of the nuclei in our present sample appear to be offset from the dynamical
centers of the galaxies as defined by either the optical continuum 
and/or the centroid of the global \HI\ profile 
(see also the Appendix). The frequent
displacement of compact nuclei from the dynamical centers of galaxies
has been discussed previously
by Miller \& Smith (1992; see also Levine \& Sparke
1998) and may offer an important clue to the formation mechanism of the nuclei.
Higher resolution spectroscopy of more extensive samples 
of these types of nuclei is
clearly desirable in order to better understand their natures, to
improve these mass estimates, and to help us understand their
dynamical relationships to the surrounding disk material.

\section{The Angular Momentum of Small Disk Galaxies\protect\label{angmom}}
In cold dark matter (CDM) galaxy formation models, protodisk baryons
tend to efficiently lose angular momentum to their dark matter halos,
resulting in a situation where the predicted sizes of galaxy
disks are too compact
by substantial factors (e.g., Navarro, Frenk, \& White 1995; Navarro \&
Steinmetz 2000).
Since our sample contains a variety of small disk galaxies, it provides an
opportunity to constrain disk angular momenta at moderate galactic masses.
These systems are of particular interest, since their 
moderate mass, yet gas-rich and relatively unevolved 
disks pose
a challenge for proposed solutions to the angular momentum problem
via internal evolution, such as intense star-forming events
(cf. Efstathiou 2000; Silk 2001).

We calculated specific angular momenta $j_D$ for the stellar disks
in our present sample using the approximate disk scale 
lengths in MG97 and assuming
$j_D\ge V_{l}/\alpha$, where the inverse exponential
scale length is $\alpha$ and $V_{l}$ is the peak of the linearly rising
rotation curve. This is a lower bound, as the \HI\ gas will have
higher specific angular momentum, as will the stellar disks of galaxies
where $V_{l}$ is approached rapidly within 1-2 scale lengths. 
The result is that we find typical angular momenta of
100-200 km s$^{-1}$ kpc for systems with $V_{l}\approx$ 100 km s$^{-1}$.
Thus this excercise
reveals that our sample galaxies appear to fall on the normal angular
momentum-rotational velocity relationship found
for giant spirals (e.g., Navarro \& Steinmetz 2000) and that our data agree with
results from \HI\ rotation curve studies in showing that disks of 
gas-rich, moderate mass galaxies have normal specific angular 
momenta (see van den Bosch et al. 2001). Our findings therefore are
consistent with the specific angular momenta 
of small, pure disk systems having been set during their
formation  (see Fall \& Efstathiou 1980),
in contradiction to the predictions of standard CDM
models (cf. Navarro \& Steinmetz 2000).

\section{Conclusions}
We have presented new optical longslit kinematic measurements for 21
low-luminosity, extreme
late-type spiral galaxies using the H$\alpha$ emission line. 
We have derived position-velocity
(P-V) curves for all 21 galaxies, as well as  deprojected
major axis rotation curves for 15 of the objects. To facilitate future
synthesis with \HI\ rotation curve measurements, we have also made all of
our P-V data available in electronic format.

Most of the
extreme late-type spirals in our sample
exhibit slowly rising, nearly linear P-V curves throughout a
significant fraction of their stellar 
disks. Many of the P-V curves continue  to rise to the
last measured point in the optical galaxy, implying  low central matter
densities. The disks of these galaxies appear to follow the
trend between specific angular momentum and rotational velocity
seen in giant spirals, suggesting that the angular momenta of disks are
not closely tied to the details of their structures or star formation 
histories, and
are intrinsically larger than predicted by current cold dark matter models.

A significant fraction of our measured optical P-V curves are not
symmetric, and our observations are consistent with $\sim$50\%
of extreme late-type spirals possessing
kinematic and/or structural asymmetries. In some cases, this may also be
manifested as an offset between the $V_{sys}$ values measured from
global \HI\ spectra and that measured 
from the optical data, although in most cases those values show
excellent agreement. We find that lopsided global \HI\
profiles appear to be a good, although not a perfect predictor of
kinematic asymmetries in extreme late-type spiral disks. 

Five of the
extreme late-type spirals in our sample possess compact, semi-stellar nuclei
that produce kinematic signatures in the galaxy rotation curves.
Rough mass estimates help to confirm that these are massive, highly
compact star
nuclei star clusters analogous to those seen in brighter galaxies. Such
nuclei appear to be a common feature of extreme late-type spiral
galaxies in spite of their apparently weak central gravitational potentials.

\acknowledgements
We are grateful to the late 
R. Schommer as well as the CTIO staff for their assistance with
the RC Spectrograph observations. LDM completed a portion of this work
while  supported by a Clay Fellowship
from the Harvard-Smithsonian Center for Astrophysics. JSG thanks 
the University of Wisconsin Graduate School 
and for their partial support of this research.


\begin{figure}
\plotfiddle{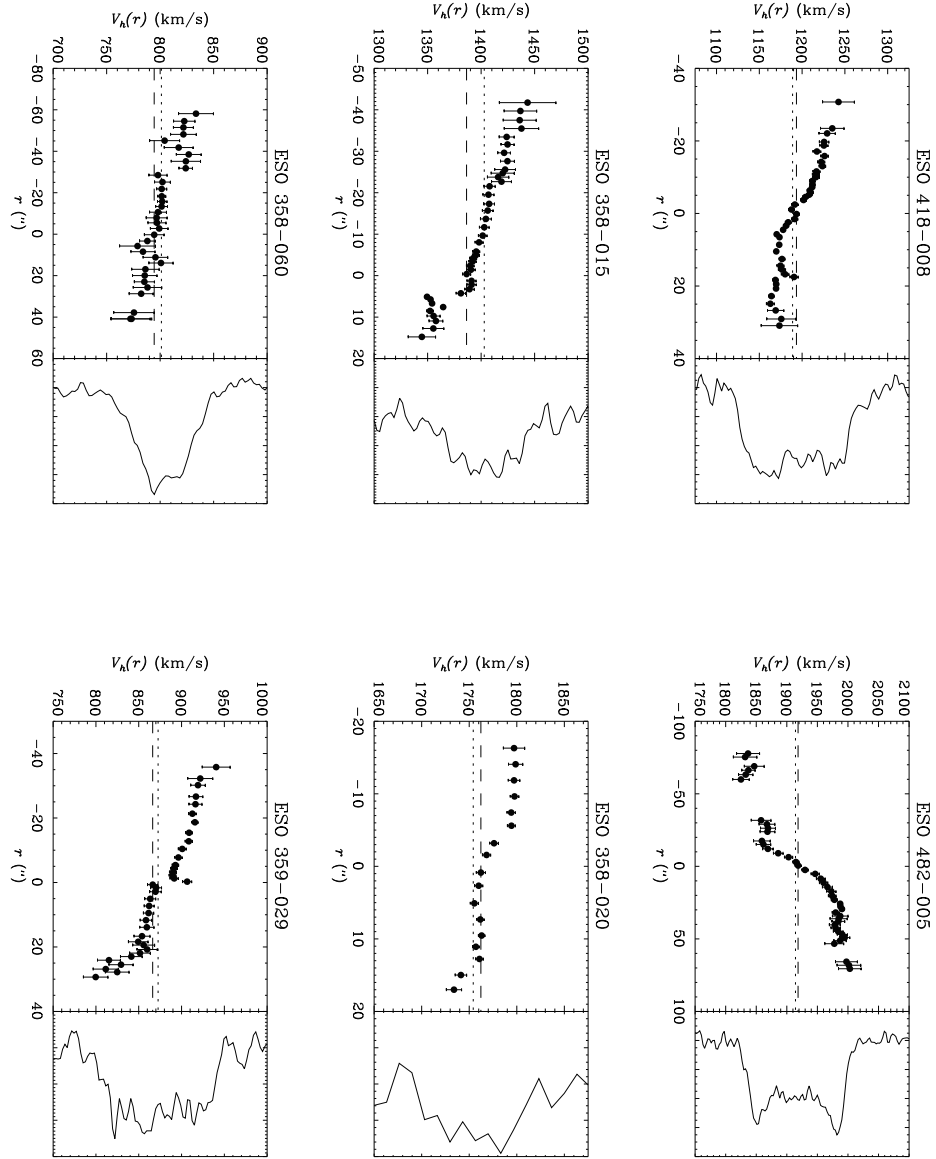}{10.0in}{0}{80}{80}{-210}{150}
\vspace{-9cm}
\caption{H$\alpha$ position-velocity (P-V) curves for the 21 extreme 
late-type
spirals observed in the present study. Axes are heliocentric radial
velocity (in kilometers per second) versus distance from the galaxy
center (as defined by the nuclear continuum), in arcseconds. When
available, the global \HI\ spectrum of the galaxy from Matthews et
al. (1998) is plotted alongside
the optical data. The global \HI\ spectra are plotted as heliocentric
velocity (on the same scale as the optical data) versus flux density (in
arbitary units). The horizontal dashed lines indicate the systemic
velocities derived from the optical observations, while the horizontal
dotted lines indicate the systemic velocities derived from the global
\HI\ observations of Matthews et al. (1998) or Fouqu\'e et
al. (1990).
\protect\label{fig:pv}}
\end{figure}

\clearpage

\begin{figure}
\plotfiddle{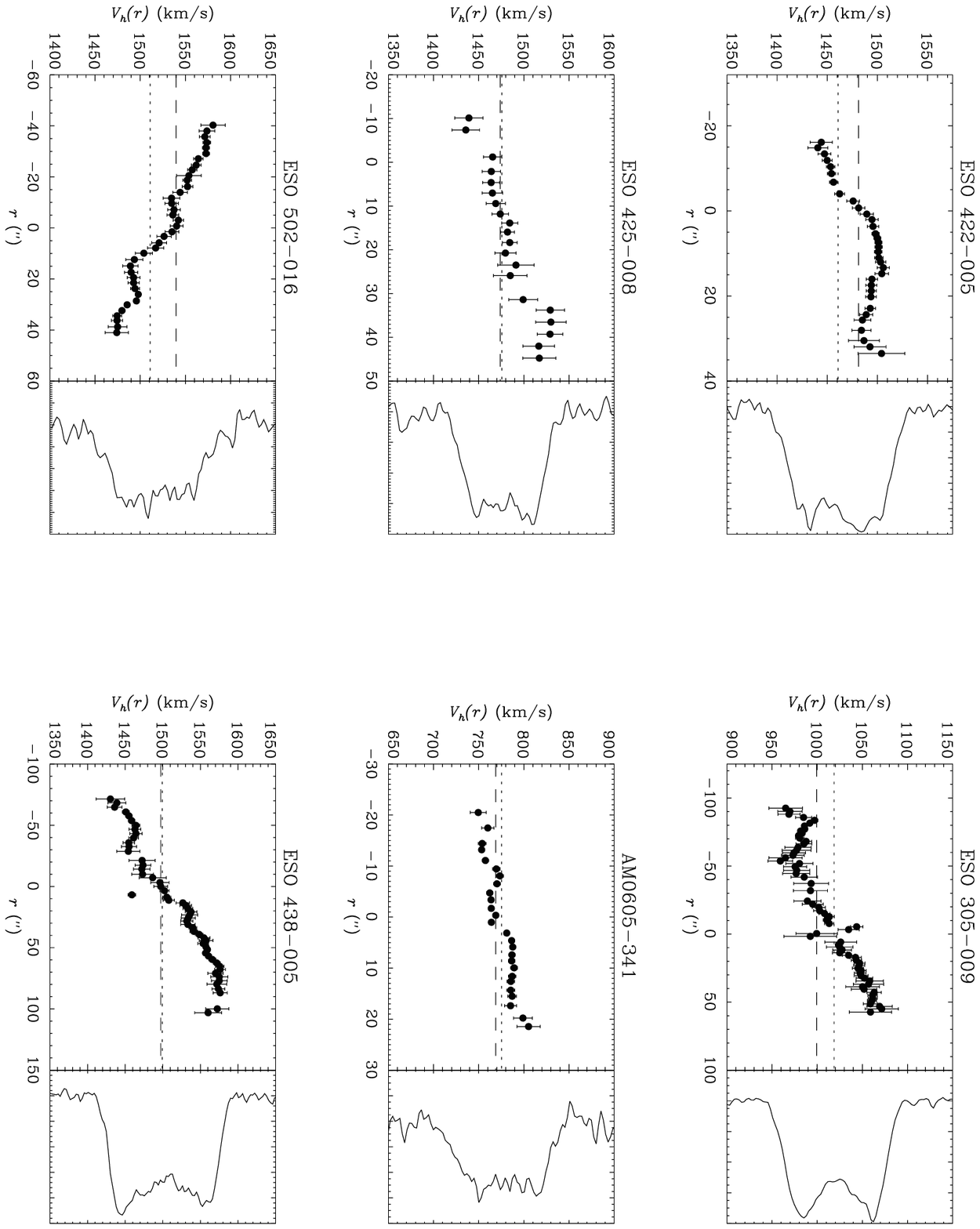}{10.0in}{0}{80}{80}{-210}{150}
\setcounter{figure}{3}
\vspace{-9cm}
\caption{cont.}
\end{figure}

\clearpage

\begin{figure}
\plotfiddle{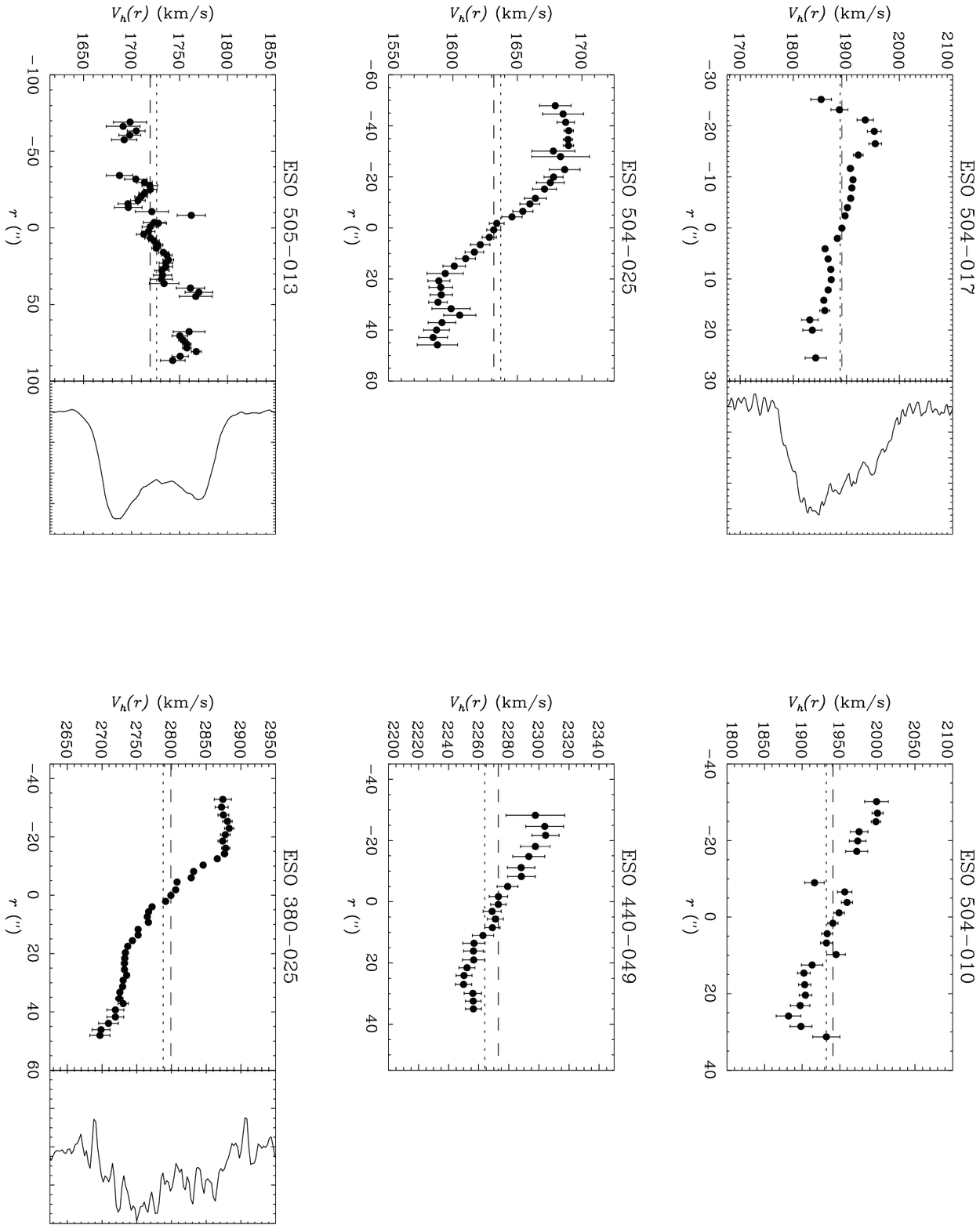}{10.0in}{0}{80}{80}{-210}{150}
\setcounter{figure}{3}
\vspace{-9cm}
\caption{cont.}
\end{figure}

\clearpage

\begin{figure}
\plotfiddle{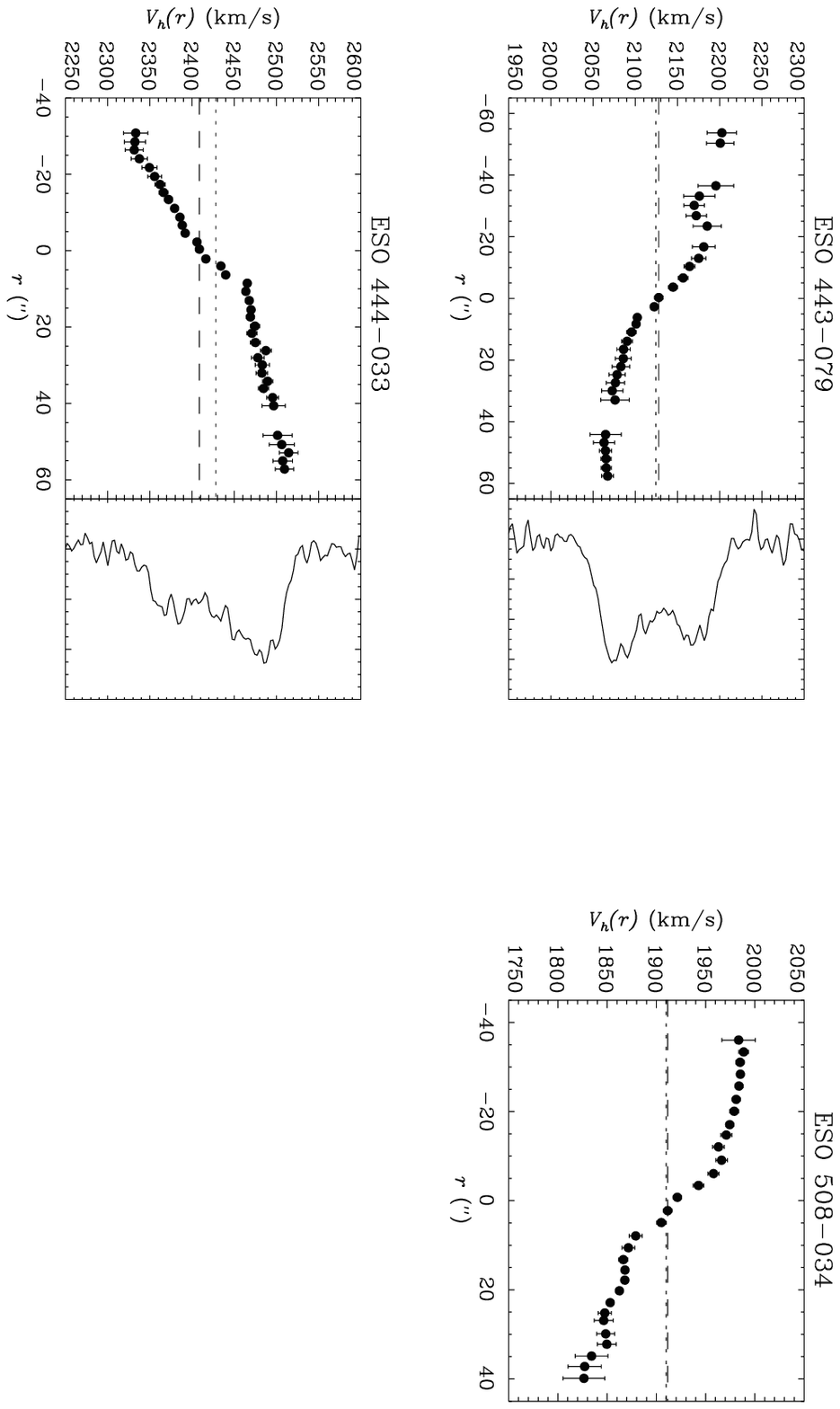}{10.0in}{0}{80}{80}{-300}{150}
\setcounter{figure}{3}
\vspace{-9cm}
\caption{cont.}
\end{figure}

\clearpage

\begin{figure}
\plotfiddle{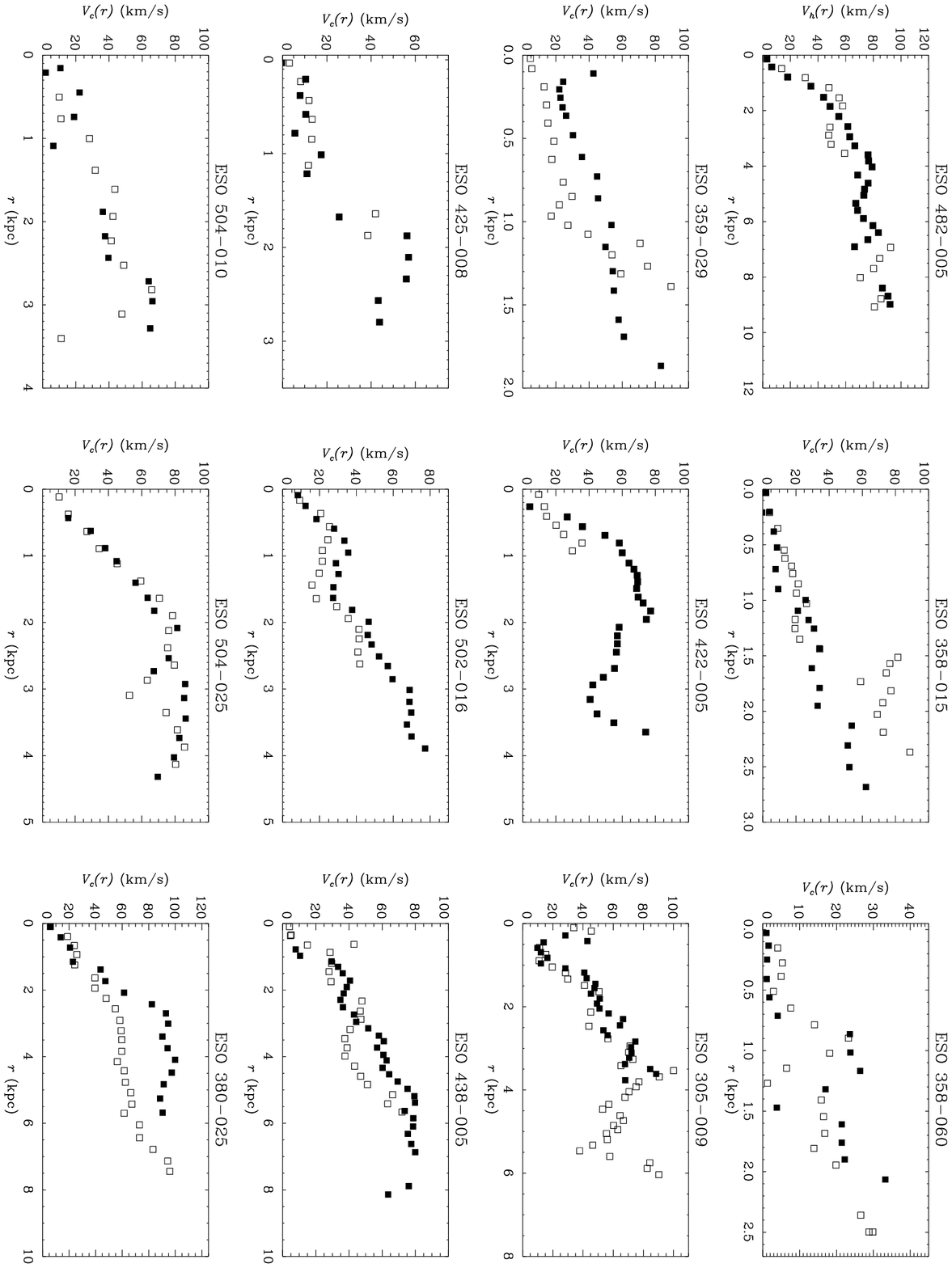}{10.0in}{0}{80}{80}{-230}{240}
\setcounter{figure}{4}
\vspace{-9cm}
\caption{Major axis rotation curves for 15 galaxies in our sample. The
data are corrected for inclination and offsets between the
spectrograph slit and the true disk major axis (see Text for further
details).  Data points from the approaching side of the galaxy are shown as
open symbols and data from the receding side are shown as 
filled symbols.\protect\label{fig:rc}}
\end{figure}

\clearpage

\begin{figure}
\plotfiddle{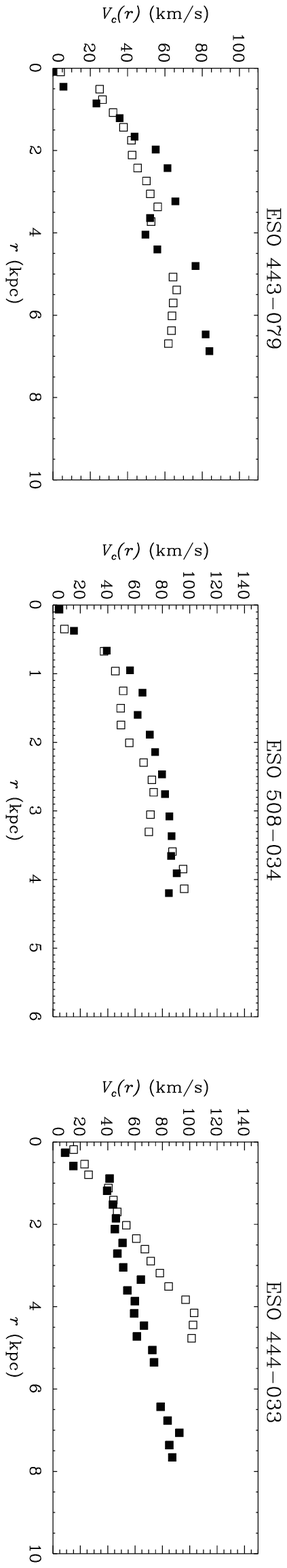}{10.0in}{0}{80}{80}{-370}{240}
\setcounter{figure}{4}
\vspace{-9cm}
\caption{cont.}
\end{figure}

\clearpage

\tablenum{1}
\begin{deluxetable}{llccccccc}
\tablecaption{Summary of Observations}
\tablehead{
\colhead{Galaxy Name}          &
\colhead{Type}  &    \colhead{$\alpha$ (1950.0)}      &
\colhead{$\delta$ (1950.0)}     &    \colhead{{\it n}} &
\colhead{$t$} & \colhead{$\phi_{ma}$} & \colhead{$\phi_{slit}$} 
& \colhead{$|\Delta\phi|$}
  \\[.2ex]
\colhead{} & \colhead{} & 
\colhead{(\phm{0}$^{h}$\phm{0}$^{m}$\phm{0}$^{s}$)} & 
\colhead{(\phm{0}$^{\circ}$\phm{0}$'$\phm{0}$''$)} & 
\colhead{} & \colhead{($s$)} &
\colhead{($^{\circ}$)} & \colhead{($^{\circ}$)} & \colhead{($^{\circ}$)} \\[.2ex]
\colhead{(1)} & \colhead{(2)} & \colhead{(3)} & \colhead{(4)} & 
\colhead{(5)} & \colhead{(6)} & \colhead{(7)} & \colhead{(8)} &
\colhead{(9)} }

\startdata

ESO418-008  &  SBdm &  03 29 28 &  -30 22 54 & 2 & 1500 & 0 & 140$^{1}$
& 40 \nl

ESO482-005 &  SBd & 03 30 52 & -24 18 06 & 2& 1800 & 79 & 80 & 1
\nl

ESO358-015 &  Sdm &  03 31 10 & -34 58 30 & 2 & 2000 & 16 & 180 & 16 \nl

ESO358-020 &  SBdm &  03 32 58 &  -32 48 18 & 2 & 1500 & 165 & 167 & 2\nl

ESO358-060 &  Sm &  03 43 18 & -35 43 30 & 2 & 2000 & 102  & 102 & 0
\nl

ESO359-029 &  Sdm & 04 10 56 & -33 07 42 &  2 & 2000 & 20 & 20 & 5
\nl

ESO422-005  & Sdm & 04 50 07 & -28 40 30 &  1 & 1000 & 35 & 15 & 20
\nl

ESO305-009 & SABc & 05 06 26 &  -38 22 30 & 2 &2000 & 90 & 63$^{1}$ &
27 \nl

ESO425-008   & SBm: & 06 04 38 & -27 52 18 & 2 & 1800 & 80 & 80 & 0\nl

AM0605-341 &SBdm &  06 05 31 & -34 11 49 & 1 & 750 & 90 & 90 & 0 \nl

ESO502-016 &  SBd & 11 02 48 & -26 21 18 &  1 & 1200 & 82 & 79 & 3\nl

ESO438-005 &  Sd & 11 06 33 &  -28 06 00 & 2 & 2000 & 62 & 60 & 2 \nl

ESO504-010 & SBdm &  11 40 32 &  -23 09 18 & 1 & 1200 & 14 & 15 & 1 \nl

ESO504-017 & Scd/BCD &  11 46 15 & -27 06 00 & 1 & 500 & 60 & 103 & 43$^{2}$
\nl

ESO504-025 &  Sd &  11 51 18 &  -27 04 18 & 2 & 1800 & 30 & 20 & 10 \nl

ESO440-049 &  Sc &   12 02 59 & -31 08 42 & 2 & 1800 & 0 & 60 & 60 \nl

ESO505-013 &  SABc &  12 03 33 & -22 34 18 &  1 & 900 & 135 & 45 
& 90$^{2}$ \nl

ESO380-025 & SBdm & 12 21 56 &  -35 07 54 & 1 & 1000 & 15 & 13 & 2 \nl

ESO443-079 &  SABd &  13 07 38 & -27 42 18 & 2 & 2000 & 0 & 180 & 0 \nl

ESO508-034   &  Spec &  13 14 13 & -25 04 24 & 1 & 900 & 132 & 129 & 3 \nl

ESO444-033 & SBdm &  13 23 16 & -31 52 12 & 1 & 1000 & 89 & 88 & 1
\nl

\enddata

\tablenotetext{1}{Slit was placed along the galaxy's bar.}
\tablenotetext{2}{Major axis position angle is 
uncertain owing to the small inclination of the disk (see Table~2).}

\tablecomments{Explanation of columns: (1) galaxy name; (2) Hubble 
classification from MG97 (3) \& (4) right 
ascension and declination, epoch 1950.0; (5) number of  exposures 
obtained; (6) total exposure time, in seconds; (7) position
angle of the disk major axis, in degrees (as measured east from north)
from Lauberts \& Valentijn 1988 or MG97; (8) position angle of 
spectrograph 
slit; (9) misalignment, in
degrees, between slit position angle and photometric major axis of galaxy.}

\end{deluxetable}

\clearpage

\begin{deluxetable}{lcccrrrcrrc}
\tablenum{2}
\tablewidth{40pc}
\tablecaption{Photometric and Spectroscopic Parameters}
\tablehead{
\colhead{Galaxy Name}          & \multicolumn{3}{c}{Photom. Props.} &
\multicolumn{3}{c}{\HI\ Props.} & \multicolumn{3}{c}{Opt. Spec. Props.} & 
\colhead{Ref.}
\\[.2ex]
\colhead{} & \colhead{$D_{26}$}  &  \colhead{$M_{V}$} &  \colhead{$i$}      &
\colhead{$W_{20}$}     &    \colhead{$W_{50}$} &
\colhead{$V_{HI}$}     & \colhead{$D_{H\alpha}$}          &
\colhead{$\Delta V_{H\alpha}$} &                        
\colhead{$V_{h_{opt}}$} & \colhead{}  \\[.2ex]
\cline{2-4} \cline{5-7} \cline{8-10}
\colhead{(1)} & \colhead{(2)} & \colhead{(3)} & \colhead{(4)} & 
\colhead{(5)} & \colhead{(6)} & \colhead{(7)} & \colhead{(8)} & 
\colhead{(9)} & \colhead{(10)} & \colhead{(11)}}

\startdata

ESO418-008  & 1.91& $-$17.4 
&  55.0 &153.5 & 125.0 & 1188.9 &0.64 & 79.6 & 1193.4 & 1\nl

ESO482-005 &  2.60 & $-$17.0 &  81.3 & 170.8 & 153.7 & 1914.2 & 1.54 & 
178.1 & 1918.4 & 1\nl

ESO358-015 &  1.53 &$-$16.3 & 46.7 & 100.2 & 67.8 & 1402.9 & 0.59 & 98.8 & 
1386.4 & 1 \nl

ESO358-020 &  1.93 &$-$17.7 & 37.1 & 117.7 & 98.1 & 1754.2 & 0.35 & 64.8 & 
1762.2 & 1\nl

ESO358-060 &   2.24 &$-$14.3 & 85.1 & 82.0 & 54.0 & 801.2 & 1.03 & 
61.2 & 794.4 & 1 \nl

ESO359-029 &  1.92 &$-$15.8 & 56.4 & 141.1 & 134.2 & 872.6 & 0.68 & 140.8 & 
866.3 & 1\nl

ESO422-005  & 1.70 & $-$16.7 & 41.7 & 126.2 & 99.0 & 1460.9 & 0.73 & 65.3 & 
1481.3 & 1\nl  
\nl

ESO305-009 &  4.63 &$-$17.2 & 52.9 & 129.8 & 113.1 & 1018.8 & 1.56 & 112.4 & 
999.4 & 1\nl

ESO425-008   & 1.20 &$-$14.1 & 78.8 & 106.5 & 91.3 & 1475.5 & 0.57 & 94.3 & 
1473.9 & 1\nl

AM0605-341 & 1.27 &$-$15.4 & 26.0 & 124.2 & 95.3 & 775.2 & 0.44 & 55.8 & 768.9 & 
1\nl

ESO502-016 &  2.46 &$-$17.2 & 67.1 & 155.0 & 102.8 & 1511.0 & 0.85 & 106.7 & 
1540.3 & 1\nl

ESO438-005 &  3.40 &$-$16.7 & 80.0 & 163.9 & 145.5 & 1499.4 & 1.86 & 145.9 & 
1497.2 & 1\nl

ESO504-010 & 1.44 &$-$16.2 & 67.2 & 145 & 127 & 1932 & 0.64 & 118.3 
& 1940.8 & 2\nl

ESO504-017 & 1.10 &$-$18.0 & 30.8 & 220.1 & 169.3 & 1888.3 & 0.55 & 123.2 & 
1891.5 & 1\nl
\nl

ESO504-025 & 2.42 &$-$17.2 & 39.9 & 137 & 118 & 1637 & 1.06 & 105.2 & 
1631.6 & 2\nl

ESO440-049 & 2.22 &$-$18.5 & 32.0 & 151 & 131 & 2264 & 0.78 & 54.5 & 
2273.0 & 2 \nl

ESO505-013 & 2.96 &$-$18.8 & 18.3 &  132.2 & 117.3 & 1726.1 & 1.62 & 82.6 & 
1719.4 & 1 \nl

ESO380-025 & 1.63 &$-$18.6 & 76.9 & 180.6 & 163.9 & 2788.1 & 0.84 & 186.1 & 
2799.2 & 1 \nl

ESO443-079 & 2.18 &$-$17.2 & 72.8 & 157.7 & 136.3 & 2124.4 & 1.16 & 132.5 & 
2127.8 & 1\nl

ESO508-034   & 1.53 &$-$17.5 & 64.6 & 167 & 150 & 1910 & 0.79 & 162.2
& 1911.5 & 2\nl

ESO444-033 & 2.03 &$-$17.9 & 74.3 & 169.0 & 147.0 & 2428.2 & 0.92 & 182.9 & 
2408.7 & 1 \nl

\enddata

\tablecomments{Photometric parameters are taken from MG97.
\HI\ parameters are taken from the references in Column~(11).
Quantities in Columns (8)-(10) are measured from the data presented 
in this work. Explanation of columns: (1) galaxy name; (2) optical
diameter in arcminutes; (3) absolute $V$ magnitude;
(4) disk inclination in degrees; (5) \& (6) measured global \HI\
profile width at 20\% and 50\% peak maximum, respectively, in
kilometers per second; (7) heliocentric recessional velocity measured from
\HI\ data, in kilometers per second; 
(8) extent of measured H$\alpha$ emission, in arcminutes; 
(9) measured amplitude of the H$\alpha$
velocity curve, in km~s$^{-1}$,
from the minimum on
approaching side to the maximum on the 
receding side of the disk (uncorrected for 
inclination and slit position angle); (10) heliocentric recessional
velocity derived from our new optical measurements (see 
Section~\ref{voptcomp}); (11) reference for 
quoted \HI\ parameters. }

\tablerefs{(1) Matthews et al. 1998; (2) Fouqu\'e et al. 1990}

\end{deluxetable}

\clearpage

\begin{deluxetable}{lccc}
\tablenum{3}
\tablewidth{20pc}
\tablecaption{Position-Velocity Data}
\tablehead{
\colhead{Galaxy Name}          & \colhead{$x$} &
\colhead{$V_{c}(x)$} & \colhead{$\sigma_{V}(x)$} 
\\[.2ex]
\colhead{} & \colhead{($''$)}  &  \colhead{(\kms)} &  \colhead{(\kms)} 
\\[.2ex]
\cline{2-4} 
\colhead{(1)} & \colhead{(2)} & \colhead{(3)} & \colhead{(4)}}

\startdata

ESO482-005  &    70.5  &   2003.0   &   18.5 \nl
            &    68.1 &    2001.5   &   19.4 \nl
            &    65.7  &   1997.5   &   17.9 \nl
            &    53.3  &   1977.6   &   15.4 \nl
            &    51.2  &     1987.1 &     10.1 \nl
             &   49.0  &   1994.7  &    8.2 \nl
             &   46.9  &    1990.7  &    8.2 \nl
             &   44.8  &    1984.0  &    7.8 \nl
             &   42.4  &    1979.8  &    7.5 \nl
             &   40.3  &    1978.5  &    9.0 \nl
             &   37.8  &   1984.2   &   10.4 \nl
             &   36.0   &  1984.8   &   14.0 \nl
             &   34.2   &   1987.3  &    12.7 \nl
             &   31.8   &   1979.9  &    5.6 \nl
              &  29.4    & 1990.1  &    3.5 \nl
              &  27.6   &   1987.6   &   3.3 \nl
              &  25.7   &   1987.3  &    3.3 \nl
              &  23.0   &   1977.8 &     5.1 \nl
              &  20.3   &   1974.2  &    6.2 \nl
              &  17.2   &  1972.8  &    7.5 \nl
             &   14.2   &   1966.4  &    8.2 \nl
             &   11.2  &   1960.1  &    6.8 \nl
              &  8.5   &   1955.5  &    7.1 \nl
              &  5.1   &   1946.4  &    6.7 \nl
              &  2.4   &  1929.8  &    5.0 \nl
              &  -0.5  &    1918.4   &   4.8 \nl
              &  -2.9   &  1914.9 &     4.9 \nl
              &  -6.3  &  1902.8  &    6.7 \nl
              &  -9.0  &   1885.9   &   7.8 \nl
              &  -12.0 &    1869.0 &      8.6 \nl
              &  -15.0 &    1861.7  &    11.8 \nl
              &  -17.5 &      1859.2  &    13.7 \nl
              &  -23.8 &    1868.3  &    12.1 \nl
              &  -26.2 &     1869.2  &    12.2 \nl
              &  -29.0   &   1867.5  &    12.9 \nl
              &  -31.7  &    1857.8   &   16.1 \nl
              &  -59.8  &    1824.9  &    13.5 \nl
              &  -63.2   &   1832.7   &   11.7 \nl
              &  -66.2  &    1837.1 &      10.9 \nl
             &   -68.9  &    1846.7  &    16.2 \nl
              &  -75.3  &   1831.8   &   19.2 \nl
              &  -77.7  &   1836.5   &   18.8 \nl

\enddata
\tablecomments{Measured H$\alpha$ position-velocity data for the 
program galaxy ESO~482-005. The data for the full sample are available
in electronic format. Columns are (1) galaxy name; (2) position $x$ along the
major axis in arcseconds; (3) heliocentric radial velocity at position
$x$ (using the
optical convention $V_{c}(x)=(\lambda-\lambda_{0})c\lambda^{-1}_{0}$, where
$\lambda$ is the measured wavelength of the H$\alpha$ emission, $c$ is
the speed of light, and $\lambda_{0}$ is the rest wavelength of
H$\alpha$); (4) uncertainty in the radial velocity measurement in \kms.} 
\end{deluxetable}

\clearpage

\appendix

\section{Comments on Individual Spectra\protect\label{sec:appendix}}
\noindent{\it ESO~418-008:} The shallow amplitude of the H$\alpha$ P-V
curve relative to the global \HI\ profile likely partially arises from the
positioning of the  spectrograph
slit along the galaxy bar, which is displaced
$\sim$40$^{\circ}$ from the galaxy major axis.

\smallskip
\noindent{\it ESO~358-015:} The break in this galaxy's P-V curve
near $V_{sys}$
appears to correspond to the location of the galaxy's semi-stellar
nucleus, which is  displaced from the dynamical center of the
galaxy as defined by both the global \HI\ profile and the optical
continuum emission. 
The H$\alpha$ P-V curve is strongly asymmetric, as is the stellar
disk of the galaxy. These strong asymmetries suggest that the offset
between the measured optical and radio
radial velocities may be real, and a consequence of a true offset between
the stellar and the \HI\ disks. Near the
galaxy's nucleus, the H$\alpha$
emission lines contain a broad wing on the redward side; some
of the H$\alpha$ emission features near the galaxy nucleus
also appear double-peaked.

\smallskip

\noindent{\it ESO~358-060:} The H$\alpha$ P-V curve extends significantly
farther on the approaching side than on the receding  side. This galaxy does
not appear to have spiral arm structure or strong internal absorption,
hence the ``wiggles''
in its P-V curve likely result from a very patchy distribution of
\HII\ regions.

\smallskip

\noindent{\it ESO~359-029:} The reversal seen near the center of this
rotation curve appears to correspond to the location of this galaxy's
semi-stellar nucleus (see also Matthews et al. 1999). Weak
asymmetries are seen in both the optical
P-V curve and the global \HI\ profile. Possible H$\alpha$ absorption
may be present.

\smallskip

\noindent{\it ESO~422-005:} It is unlikely that a 
misalignment of 20$^{\circ}$
between the spectrograph slit and the disk major axis can
account for the discrepancy between the optical and radio $V_{sys}$
values, or the odd shape of the optical P-V curve, which on the receding
side shows a rise, then
falls to near $V_{sys}$ before once again rising. This galaxy
shows an asymmetry in the global \HI\ profile 
in terms of the shape of the rotation peaks, as well as a
weak asymmetry in the
stellar disk (see also MG97).

\smallskip

\nin{\it ESO~305-009:} Strong nuclear continuum present.
A reversal  in the optical P-V curve near
$V_{sys}$ occurs at the
location of this galaxy's semi-stellar nucleus.  A misalignment of
$\sim$27$^{\circ}$
between the slit placement and the galaxy major axis may partially
account for the odd shape of the rotation curve, which on the
approaching side shows a rise, then
falls to near $V_{sys}$ before once again rising.

\smallskip

\nin{\it ESO~425-008:} The peak of the optical continuum shows
a large spatial asymmetry relative to the velocity centroids of both the
optical P-V curve and the global \HI\ profile.

\smallskip

\nin{\it AM~0605-341:} This galaxy shows very little rotation in
its stellar disk.
The global \HI\ profile is significantly broader than the amplitude of
the H$\alpha$ P-V curve, indicating this galaxy must have extended
\HI\ and reach its maximum rotational velocity well outside of its
stellar disk. The H$\alpha$
lines near the center of this galaxy are rather broad. Our spectra suggest
presence of a possible
starburst nucleus.

\smallskip

\nin{\it ESO~502-016:}  The asymmetries in the stellar disk and
the global \HI\ profile 
of this galaxy suggest that the observed
offset between the optical and radio $V_{sys}$ values may be real. The
galaxy does not show any obvious features in the optical disk to
explain the  wavy appearance of the P-V curve.

\smallskip

\nin{\it ESO~438-005:} A small break occurs in the H$\alpha$ P-V curve
of this galaxy near
$V_{sys}$, which may result from this galaxy's weak bar.

\smallskip

\nin{\it ESO~504-010:} The breaks that occur in the H$\alpha$ P-V
curve, slightly offset from $V_{sys}$, may be 
caused by this galaxy's bar.

\smallskip

\nin{\it ESO~504-017:} The slit was misaligned by
$\sim$43$^{\circ}$ from the galaxy major axis during the observations
of this galaxy. However, since the disk has a relatively low inclination
($i\approx$31$^{\circ}$), this probably did not affect the
observed P-V curve significantly, which shows a very low amplitude in the inner
regions of the galaxy. This galaxy was shown by Matthews et al.
(1998) to have rather extended \HI.

\smallskip

\nin{\it ESO~504-025:} The slight ``glitch'' seen in the optical P-V
curve near $V_{sys}$ appears to be due to this galaxy's semi-stellar nucleus.
The P-V curve is observed to flatten off, but distinct
dips are seen on the flat part of the P-V curve near radii of $\pm33''$
from the galaxy center.

\smallskip

\nin{\it ESO~440-049:} Very strong nuclear continuum present.
The bright continuum as well as the slight perturbation seen near $V_{sys}$
may be due to this galaxy's small bulge component.

\smallskip

\nin{\it ESO~505-013:} Strong nuclear continuum present.
This galaxy contains spiral arms clearly
delineated by \HII\ regions, as well as a semi-stellar nucleus. The
ripples in the optical P-V curve may be due to the spiral arm pattern,
while the rotation curve reversal seen near $V_{sys}$ is likely due to
the semi-stellar nucleus. The rotation curve is spatially more extended
on
the receding side, while the global \HI\ profile contains slightly 
more flux on
the approaching side. The slit was placed near the galaxy's minor axis
during these observations, but because of the galaxy's small
inclination ($i\approx18^{\circ}$), the observed shape of the
P-V curve was probably not
significantly affected.

\smallskip
\nin{\it ESO~380-025:} The shape of the H$\alpha$ P-V curve shows
significant differences on the receding and approaching sides, with the
receding side flattening off, and the approaching side continuing to
rise to the last measured point.

\smallskip

\nin{\it ESO~443-079:} The H$\alpha$ P-V curve appears to flatten near
2.2 radial scale lengths. Both the approaching and receding sides show
a break near at roughly $\pm $40$''$ from the galaxy center.

\smallskip

\nin{\it ESO~444-033:} Both the approaching and receding sides of
the H$\alpha$ P-V curve are quite linear in shape, but have differing slopes.
The $V_{sys}$ value measured from the global \HI\ spectrum appears to
better correspond to the rotation  curve center than the
optically-derived value. Both the optical galaxy and the global \HI\
profile are very asymmetric, thus the observed offset between the
optical and radio $V_{sys}$ values may be real.

\end{document}